\documentstyle[12pt]{article}

\begin{document}

\pagestyle{empty}

\hfill hep-th/9610074

\hfill CERN-TH/96-277

\hfill USC-96/HEP-B6

\bigskip \bigskip \bigskip \bigskip

\begin{center}
{\LARGE BLACK\ HOLE\ ENTROPY\medskip \medskip \\[0pt]
REVEALS\ A\ 12th\ DIMENSION}\footnote{{ Research supported by the U.S.
Department of Energy under grant number DE-FG03-84ER40168}}
\bigskip \bigskip \bigskip \\[0pt]

{\Large Itzhak Bars}\footnote{ On sabbatical leave from the
Department of Physics and Astronomy, University of Southern California, Los
Angeles, CA 90089-0484, USA.}{ \bigskip \bigskip \\[0pt]}

{\large TH Division, CERN, CH-1211 Geneva 23, Switzerland}

\medskip \bigskip \bigskip \bigskip \bigskip

{\bf ABSTRACT}
\end{center}

\noindent
{The Beckenstein-Hawking black hole entropy in string theory and its
generalizations, as expressed in terms of charges that correspond to central
extensions of the supersymmetry algebra, has more symmetries than U-duality.
It is invariant under transformations of the charges, involving a 12th (or
13th) ``dimension''. This is an indication that the secret theory behind
string theory has a superalgebra involving Lorentz non-scalar extensions
(that are not strictly central), as suggested in S-theory, and which could
be hidden in M- or F- theories. It is suggested that the idea of spacetime
is broader than usual, and that a larger ``spacetime" is partially present
in black holes.}

\bigskip \bigskip

\noindent CERN-TH/96-277

\noindent {October 1996}

\vfill\eject

\setcounter{page}1\pagestyle{plain}

\section{Introduction}

In recent work in string theory, and its extensions to a secret
M-,F-,S-,Y-,... theory, a microscopic explanation of the Beckenstein-Hawking
area-entropy law for string-related black holes is obtained by counting \cite
{st-vaf}-\cite{kal-raj} the number of D-brane states \cite{pol}. For
(nearly) extreme black holes the entropy is given in terms of the charges
carried by these states \cite{kal}-\cite{kal-raj}. These charges correspond
to central extensions of the extended supersymmetry algebra, and it has been
shown that the entropy is an invariant under U-duality \cite{ht}
transformations applied on these charges. In particular, for 4-dimensional
black holes the entropy is given by the quartic invariant of $E_{7\left(
7\right) }$ \cite{kal} and for 5-dimensional black holes it is given by the
cubic invariant of $E_{6\left( 6\right) }$ \cite{kal-fer}\cite{ver}\cite{hc}
. In higher dimensions the entropy for non-extremal black holes has also
been given as an invariant of the U-duality group \cite{kal-raj}\cite{hc}. 

The entropy, which is related to just counting states, is really independent
of the moduli, although it is expressed in terms of central extensions that
depend on the moduli \footnote{
Define $K$ as the largest compact subgroup in the duality group $U.$ The
moduli are classifiedas the right-coset $U/K.$ Global $U$ transformations,
applied from the left on the moduli, and on quantized charges, induce
moduli-dependent $K$ transformations that are applied on the supercharges as
well as the maximal extensions of the superalgebra. The supercharges as well
as the moduli-dependent central extensions are singlets of $U,$ but
transform under this induced $K$, which is continuous, since it depends on
the moduli. The classification of these operators under $K$ is tabulated in 
\cite{ibbeyond} for various dimensions. The black hole entropy is expressed
in terms of central extensions that are moduli dependent, but the moduli
dependence can be stripped away under $U$ transformations since the entropy
is really independent of the moduli.}. This simple fact distinguishes the
entropy as an ideal quantity for exploring the symmetries of the secret
theory. If the underlying secret theory has a dynamical symmetry structure,
such that states or operators fall into reprentations of its algebra, then
the entropy simply counts the multiplicity in a way consistent with the
representations of the dynamical symmetry. If the secret theory is suspected
to have some hidden symmetries, {\it they should show up as invariances of
the entropy}. For example, the U-duality invariance of the entropy, for all
values of the moduli, as if U is continuous rather than discete, is
explained through the reasoning given in this paragraph.

Furthermore, if there are suspected extra ``dimensions'', the entropy should
be invariant under transformations that mix the extra `dimensions'' with the
10 dimensions that are explicit in string theory. One of the approaches to
the secret theory behind string theory is S-theory \cite{ibstheory}, which
concentrates on the global dynamical symmetry aspects of the secret theory.
S-theory predicts a number of symmetries, some of which are outside of
U-duality and which involve a 12th and 13th ``dimension''. It is the goal of
this paper to demonstrate that the entropy given below is indeed invariant
under transformations that mix the hidden ``dimensions'' with the 10 or 11
``familiar'' ones. Hence the entropy serves to show the presence of extra
``dimensions'' (where ``dimension'' is used in the sense explained in \cite
{ibstheory} and in the footnote \footnote{
As in \cite{ibstheory}, it must be emphasized that the 12th (or
13th)``dimension'' is not to be understood as a naive extension of the other
11 dimensions. Rather, it is the spinor space of 12D (or projected 13D
spinor space) that is relevant, with 32 Weyl-Majorana supercharges $Q_\alpha 
$ of A or B type. The supercharges close into $\left\{ Q_\alpha ,Q_\beta
\right\}=S_{\alpha \beta }\, (A\,\,or\,\,B\,\,types)$ where the 32x32
symmetric matrix $S\,$ contains 528 bosons. The 11 usual momenta are members
of $S$, and they are on an equal footing with the 528 bosons. In this paper
12D (or 13D) is used in this sense, i.e. involving the spinors in these
``dimensions'', and arranging the 528 bosons into subsets consistent with
12D Lorentz transformations $SO(10,2)$ (the 13th ``dimension'' is mixed by $
A\Leftrightarrow B$ duality \cite{ibstheory}). To emphasize this point, in
this paper the word ``dimension'' will appear in quotes whenever appropriate.
}).

Specifically, in 4D define the N=8 superalgebra and its central extensions
in an SU(8) basis 
\begin{eqnarray}
\left\{ Q_{\alpha A},Q_{\dot{\beta}\bar{B}}\right\} &=&\sigma _{\alpha \dot{
\beta}}^\mu \,P_\mu \,\,\delta _{A\bar{B}}  \nonumber \\
\left\{ Q_{\alpha A},Q_{\beta B}\right\} &=&\left( i\sigma _2\right)_{\alpha
\beta }\,Z_{AB},\quad \left\{ Q_{\dot{\alpha}\bar{A}},Q_{\dot{\beta}\bar{B}
}\right\} =\left( i\sigma _2\right) _{\dot{\alpha}\dot{\beta}}\,Z_{\bar{A}
\bar{B}}^{*},  \label{4dz}
\end{eqnarray}
where $\mu =0,1,2,3$ are SO(3,1) vector indices, $\alpha ,\dot{\alpha}=1,2$
are Weyl spinor indices, and $A,\bar{A}=1,\cdots ,8$ are {\bf 8,8}$^{*}${\bf 
\ } indices of SU(8). Then {\bf \ }the extremal black hole entropy $
S=A_4/4G_4=2\pi \sqrt{I_4}\,\,$ is given by the quartic invariant of $
E_{7(7)} $ \cite{kal} 
\begin{equation}
I_4=Tr\left( ZZ^{*}ZZ^{*} \right) -\frac 14\left[ Tr\left(
ZZ^{*}\right)\right] ^2+4Pf \left( Z\right) +4Pf\left( Z^{*}\right) ,
\label{ent4}
\end{equation}
where $Pf\left( Z\right) = \varepsilon ^{A_1\cdots
A_8}Z_{A_1A_2}Z_{A_3A_4}Z_{A_5A_6}Z_{A_7A_8}/ \left( 4!2^4\right) .$ In this
case the central charges $Z,Z^{*}$ are antisymmetric matrices that form the
representations {\bf 28,28}$^{*}$ of $SU(8)$. The 56 real components of
these central charges depend on 56 quantized charges that form the {\bf 56}
of $E_{7(7)}$ as well as on the 70 moduli classified as the coset of $
E_{7(7)}/SU(8).$ Because of the invariance under $E_{7(7)}$ the dependence
on the 70 moduli can be stripped away, leaving only the dependence on the 56
quantized charges. On the other hand, the mass of the black hole is not an
invariant under $E_{7(7)}$, and therefore it does depend on the 70 moduli
and 56 quantized charges. It is, of course, an invariant under under $SU(8)$
which is the explicit symmetry of the superalgebra. 

Similarly, in 5D define
the $N=8$ superalgebra in an $Sp(8)$ basis 
\begin{equation}
\left\{ Q_{\alpha A},Q_{\beta B}\right\} = P_\mu \left( C\gamma ^\mu
\right)_{\alpha \beta }\Omega _{AB}+ C_{\alpha \beta }\,\,Y_{AB}  \label{5dz}
\end{equation}
where $\mu =0,1,2,3,4$ are $SO(4,1)$ vector indices, $\alpha =1,2,3,4$ are
Dirac spinor indices, and $A=1,\cdots ,8$ are indices in the fundamental
representation of $Sp(8)$. Then the extremal black hole entropy $
S=A_5/4G_5=2\pi \sqrt{I_5}$ is given by the cubic invariant of $
E_{6\left(6\right) }$ \cite{kal-fer} \cite{ver} \cite{hc}. 
\begin{equation}
I_5=Tr\left( Y\Omega Y\Omega Y\Omega \right) .  \label{ent5}
\end{equation}
In this case $\Omega _{AB}$ is the one-component antisymmetric invariant of $
Sp(8)$, while $Y_{AB}$ is an antisymmetric matrix, orthogonal to $
\Omega_{AB},$ Tr$\left( Y\Omega \right) =0,$ that forms the {\bf 
27\thinspace\thinspace }of $Sp(8).$ The 27 central extensions $Y_{AB}$
depend on 27 quantized charges that form the {\bf 27 }of $E_{6\left(
6\right) }$ as well as on the 27 moduli classified as the coset $E_{6\left(
6\right) }/Sp(8).$ In higher dimensions, as well as in dimensions 4,5, the
nearly extremal black hole entropy is also expressed in terms of the central
extensions of the corresponding superalgebra \cite{kal-raj} \cite{hc}. Our
discussion below generalizes to all cases.

The strategy in this paper is to first clarify the (broken) symmetries that
involve the 12th (or 13th) ``dimension'', rewrite the central extensions
above in a basis appropriate to these symmetries, and finally show that
indeed the entropy, as well as the mass of the black hole are invariants
under transformations that include the 12th (or 13th) dimension. 

One implication of our discussion is that the concept of ``spacetime'' is
broader than usually envisaged. Through the entropy considerations discussed
in this paper it is demonstrated that the spacetime superalgebra contains
more than the usual spacetime momenta plus Lorentz scalar central extensions
which have become familiar by now. A larger space that includes {\it Lorentz
non-scalar extensions} (which are not strictly central) is shown to be
relevant to fully realize the minimal symmetry consistent with 12D (or 13D)
that is present in the black hole entropy. This suggests the presence of
additional moduli. The broader concept of spacetime is further discussed in
the last section of this paper.

\section{Mixing ordinary and extra ``dimensions''}

\subsection{11th dimension}

First consider a familiar case in a framework that will generalize. From the
M-theory point of view \cite{witten}-\cite{vafaf} the black hole entropy
should reflect the presence of an 11th dimension. This manifests itself in
the entropy by its invariance under transformations that mix the 11th
dimension with the compactified ones. These transformations are applied on
the central extensions which form multiplets. 

Let's identify the symmetry
when M-theory is toroidaly compactified down to $d$ dimensions. In the
superalgebra of M-theory, one of the central extensions is the 11th
momentum. The Kaluza-Klein momenta for $c$ compactified dimensions of string
theory are also viewed as $c$ central extensions of the superalgebra. The
superalgebra is covariant under the rotations 
\begin{equation}
SO(c+1)
\end{equation}
that mix the $c$ compactified dimensions of string theory with the 11th
dimension of M-theory (note that $d+c+1=11)$. The superalgebra has more
central extensions besides the Kaluza-Klein momenta, such as winding numbers
and others. All central extensions fall into multiplets of $SO(c+1)$, as can
be seen by starting from the 11D superalgebra with all possible central
extensions \cite{townsend}, and then compactifying \cite{ib11}\cite{ibbeyond}
. In $d=4$ the 56 central extensions form the following $SO(7)$ multiplets:
(7+21+21+7). Similarly, in $d=5$ the 27 central extensions form the
following $SO(6)$ multiplets (6+15+6). 

$SO(c+1)$ is one of the subgroups of the U-duality group 
\begin{equation}
U\supset K\supset SO(c+1),
\end{equation}
where $K$ is the largest compact subgroup of $U$. In $d=4$ one has $K=SU(8)$
 and $SO(c+1)=SO(7).$ In $d=5$ one has $K=Sp(8)$ and $SO(c+1)=SO(6).$ The
presence of the extra 11th dimension in M-theory can be seen in all
expressions by covariance under $SO(c+1).$ This is a consequence of the
covariance of the superalgebra under $SO(c+1),$ which in turn requires
multiplets of $SO(c+1)$ for all operators and all states. Therefore, the
black hole entropy must be invariant under $SO(c+1).$ 

In fact, the black hole entropy written above is invariant under the much
larger $U$ transformations, hence it is automatically invariant under its
(continuous) subgroups $K\supset SO(c+1).$ This is consistent with the
presence of the 11th dimension on the same footing with the other $c$
compactified dimensions. In this example $SO(c+1)$ invariance is a trivial
consequence of $U$-invariance. This is not true for the 12th (or 13th)
``dimension'', but a reasoning similar to the one given above for the 11th
dimension will apply.

\subsection{12th ``dimension''}

The compactified (A or B type) superalgebra has larger isometries than $
SO(c+1).$ The Lorentz isometry $SO(d-1,1)$ in flat spacetime is evident 
\footnote{
It has been explained (see e.g. \cite{hor}) that string BPS states in {\it 
flat} spacetime evolve into black holes in {\it curved spacetime} as the
coupling becomes stronger. Therefore, the counting of black hole states are
in one to one correspondance with the counting of BPS states in flat
spacetime.}. Here we are concerned with the isometries that commute with the
Lorentz group. The structure and interrelationships among the internal
symmetries and dualities is explained in \cite{ibbeyond}\cite{ibstheory}
through the following diagram 
\begin{equation}
\begin{array}{l}
\left. 
\begin{array}{c}
\left. 
\begin{array}{c}
{\ c\,\,} {\rm {\ compact\,\,+}} \\ 
{\ 2\,\,\,} {\rm {\ hidden\,dims}}
\end{array}
\right. \\ 
SO(c+1,1) \\ 
\downarrow
\end{array}
\right. \quad \rightarrow \quad \,\,\,\,\,\,\left. 
\begin{array}{c}
{\rm {\ two\,classifications}} \\ 
{\rm {\ of\,generators\,\,\&\,\,states:}} \\ 
{\rm under\,\, } SO(c+1,1)\, {\rm \, or\,under\,\,} {\ K}
\end{array}
\right. \\ 
\left. 
\begin{array}{l}
\left. 
\begin{array}{l}
SO(c+1),\, \, 1\,\,{\rm hidden\,\, dim .} \\ 
SO(c)\,\,{\rm is\,\,common\,\,subgroup} \\ 
SO(c)_L\otimes SO(c)_R
\end{array}
\right\} \rightarrow \left. 
\begin{array}{c}
\uparrow \\ 
K \\ 
{\rm {\ max.\,\,compact\,\,in\,\,}}U
\end{array}
\right. \\ 
\begin{array}{l}
\,\,\,\,\,\,\,\,\,\,\,\,\,\,\,\,\,\,\,\,\,\,\,\,\,\,\,\, \uparrow \\ 
T\,-\, {\rm duality}\,\,SO(c,c)
\end{array}
\end{array}
\right\} \rightarrow 
\begin{array}{l}
U \\ 
{\rm duality}
\end{array}
\end{array}
\end{equation}

The largest isometry that is described by a compact group is $K.$ It
contains not only $SO(c+1)$ but also $SO(c)_L\otimes SO(c)_R$ (which is the
maximal compact group in $SO(c,c)$ of T-duality ). The largest isometry
described by a non-compact group (also commuting with the Lorentz group) is $
C\equiv SO(c+1,1)$ (see below for more details). The extra spacelike 11th
dimension and timelike 12th ``dimension'' are included in this isometry. The
Lorentz group together with $C$ are contained in $SO(10,2)$ which mixes all
the ``dimensions'' 
\begin{equation}
SO(d-1,1)\times SO(c+1,1)\subset SO(10,2).
\end{equation}
For example, 
\begin{eqnarray*}
{\rm {\,}}d &=&4,\,\,c=6\,\,:{\rm {\,\,}}K=SU(8)\,\,\,\,\,\,({\rm {in\,\,\,}}
U=E_{7\left( 7\right) }),\,\,\,{\rm {and\,\,\,}}C=SO(7,1), \\
{\rm {\,}}d &=&5,\,\,c=5\,\,:\,K=USp(8)\,\,\,({\rm {in\,\,\,}}U=E_{6\left(
6\right) }),\,\,\,{\rm {and\,\,\,}}C=SO(6,1).
\end{eqnarray*}
Even though $U$ is non-compact, generally it does not contain $C$ \footnote{
For example, in $d=8,\,c=2,$ one has $U=SL(3)\otimes SL(2)$ and $C=SO(3,1)$.
 Clearly $C$ is not contained in $U$ because it is not possible to match the
compact and non-compact generators. Similarly, in $d=5$ the group $
E_{6\left( 6\right) }$ does not contain $SO(6,1)$. In some dimensions $U $
is large enough to contain a subgroup similar to $C$. For example, in $d=9$, 
$c=1,$ one has $U=SL(2)\otimes SO(1,1)$, and $C=SO(2,1)$ is isomorphic to $
SL(2,R)$. Similarly, in $d=4$ the group $E_{7\left( 7\right) }$ contains an $
SO(7,1)$ isomorphic to $C$. But it seems that such subgroups, when they
exist, are different than $C$ since the latter acts also on the momentum $
P_\mu$, as explained in the text.}. The essential difference between $K$ and 
$C$ is described as follows. For $d\leq 11$ , $K$ acts as an internal
symmetry group under which the $d$-dimensional momentum $P^\mu $ is a
singlet (see below). However, under the group $SO(c+1,1)$ the momentum $
P^\mu $ is not a singlet, it mixes with other extensions that have a Lorentz
index. This is related to the fact that in a 12D $SO(10,2)$ covariant
notation the A-type superalgebra 
\begin{eqnarray}
&&\left\{ {Q_\alpha ,Q_\beta }\right\} =S_{\alpha \beta }  \nonumber \\
&&S=\frac{1+\gamma _{13}}2C\left( \gamma ^{M_1M_2}\,\,Z_{M_1M_2}\,+\gamma
^{M_1\cdots M_6}\,\,\,Z_{M_1\cdots M_6}^{+}\right)   \label{12d}
\end{eqnarray}
contains no 12D momentum operator $P_M$; a property that prevents the
appearance of two time coordinates, or two time translation operators,
labelled by $M=0,0^{\prime }$ \cite{ibbeyond}\cite{ibstheory} (the same is
true for the B-type superalgebra). The time translation operator is $
Z_{00^{\prime }}.$ In compactifications to lower dimensions, the 12D
antisymmetric tensor $Z_{M_1M_2}$ yields the momentum operator in lower
dimensions in the form $P_\mu \equiv Z_{\mu 0^{\prime }}$. Under the
transformations generated by $C=SO(c+1,1)$ this momentum mixes with the
other members of $Z_{\mu m}$ (defined in (\ref{comp}) below). Thus, $C$ is
clearly not in $U.$ 

The common largest compact subgroup of $K$ and $C$ is $SO(c+1)$ 
\begin{equation}
K\cap C=SO(c+1).
\end{equation}
The black hole entropy in every dimension $d$ is invariant under $U\supset
K\supset SO(c+1)$ and therefore reveals the presence of the extra 11th
dimension, as discussed above. Our aim is to show that it is also invariant
under $C=SO(c+1,1)$ and therefore indicates the presence of a 12th
``dimension'' as well. As explained above this does not automatically follow
from the known $U$ invariant expressions of the entropy since generally 
\begin{equation}
C\,\,{\rm is\,not\,in\,\,}\,U.
\end{equation}

Nevertheless, the black hole entropy must be invariant under $C$. The
mathematical reason behind this expectation is as follows. The (nearly)
extreme black hole states are completely specified by their charges. These
charges are classified by either $K$ or $C. $ As indicated on the diagram
above, since both $K$ and $C$ are isometries of the superalgebra of the
secret theory it must be possible to classify the operators and states under
either $K$ or $C.$ Completeness of states is equivalent to saying that sums
of $K$-representations are expressible as sums of $C$-representations. Since
the black hole entropy just counts BPS states (that fit into the short
representations of the superalgebra) it must be invariant under the
relabelling of the superalgebra under either $K$ or $C.$ Since the counting
of the total number of states cannot depend on how they are classified, the
entropy must be invariant under $K$ as well as $C.$

In order to explain this symmetry we must deal with the fact that the
momentum appears to be non-invariant under $C;$ a new definition consistent
with $C$-invariance will be introduced.

\section{Classification under $C$}

Under compactification the $SO(10,2)$ spinor $\alpha $ and vector indices $M$
of eq.(\ref{12d}) are reclassified under $SO(d-1,1)\otimes SO(c+1,1)$ as
follows 
\begin{eqnarray}
Q_\alpha &\rightarrow &Q_{\alpha a}{\rm {(spinor\,\,in\,\,}}d\,\,{\rm {dims}}
\times {\rm {spinor\,\,in\,\,}}c+2\,\,{\rm {dims)}}  \nonumber \\
M &\rightarrow &\mu \,\,(d\,\,{\rm {dims}})\oplus \,m\,(c+2\,\,{\rm {dims.)}}
\nonumber \\
Z_{M_1M_2} &\rightarrow &Z_{\mu _1\mu _2}\oplus Z_{\mu _1m_2}\oplus
Z_{m_1m_2}  \label{comp} \\
Z_{M_1\cdots M_6}^{+} &\rightarrow & {\rm {similar,\,but\,\,impose\,\,self\,
\,duality\,\,in\,\,12D.}}  \nonumber \\
\gamma _{M=\mu } &\rightarrow &\gamma _\mu \otimes \gamma _{c+3}  \nonumber
\\
\gamma _{M=m} &\rightarrow &\gamma _{d+1}\otimes \gamma _m  \nonumber
\end{eqnarray}
where $\gamma _{d+1}$ and $\gamma _{c+3}$ are the analogs of $\gamma _5$ in
4D. This assigns the correct transformation properties under $C$ to all
operators. For clarity, let's write out explicitly the general form of the
superalgebra in 4D and 5D.

\subsection{4D superalgebra}

For $d=4,\,c=6,$ the 32 real supercharges are relabelled in the form of a
complex $Q_{\alpha a}$ and $Q_{\dot{\alpha}\dot{a}}=Q_{\alpha a}^{\dagger },$
where 
\begin{eqnarray}
\alpha  &:&(1/2,0),\,\,\dot{\alpha}:(0,1/2){\rm {\,\,\,\,\,\,\,of\,\,\,\,}}
SO(3,1) \\
a &:&{\bf 8}_{+},\quad \dot{a}:{\bf 8}_{-}\quad {\rm {of\,\,\,\,}}SO(7,1) 
\nonumber
\end{eqnarray}
Note that the $SO(7,1)$ Weyl spinor indices $\left( a,\dot{a}\right) $ are
different than the $SU(8)$ indices $\left( A,\bar{A}\right) $ used in eq. (
\ref{4dz}), although they have the same number. The maximally extended
superalgebra with 528 bosonic generators takes the form 
\begin{eqnarray}
\left\{ Q_{\alpha a},Q_{\beta b}\right\}  &=&\left( i\sigma _2\right)
_{\alpha \beta }\,z_{ab}+\left( i\sigma _2\vec{\sigma}\right) _{\alpha \beta
}\,\cdot \vec{F}_{ab}  \nonumber \\
\left\{ Q_{\dot{\alpha}\dot{a}},Q_{\dot{\beta}\dot{b}}\right\}  &=&\left(
i\sigma _2\right) _{\dot{\alpha}\dot{\beta}}\,z_{\dot{a}\dot{b}}^{*}+\left(
i\sigma _2\vec{\sigma}\right) _{\dot{\alpha}\dot{\beta}}\,\cdot \vec{F}_{
\dot{a}\dot{b}}^{*}  \label{4d} \\
\left\{ Q_{\alpha a},Q_{\dot{\beta}\dot{b}}\right\}  &=&\sigma _{\alpha \dot{
\beta}}^\mu \,\left( \gamma _{a\dot{b}}^m\,P_{\mu m}+\gamma _{a\dot{b}
}^{m_1m_2m_3}A_{\mu m_1m_2m_3}\right)   \nonumber
\end{eqnarray}
Here the 8$\times $8 $SO(7,1)$ hermitian gamma matrices $\gamma ^m=(1,\gamma
^k)$ in Weyl spinor space are analogous to the Pauli matrix representation
of the $SO(3,1)$ hermitian gamma matrices $\sigma ^\mu =\left( 1,\vec{\sigma}
\right) $ applied on Weyl spinors$.$ 
\begin{eqnarray}
SO(3,1),\,\,\,Weyl &:&\quad \sigma ^\mu =\left( 1,\vec{\sigma}\right)  \\
SO(7,1),\,\,\,Weyl &:&\quad \gamma ^m=(1,\gamma ^k)  \label{gam8}
\end{eqnarray}
The seven $\gamma ^k$ are hermitian, purely imaginary and antisymmetric. The 
$SO(3,1)\times SO(7,1)$ classification of the 528 real operators on the
right hand side, and the number of their real components are as follows 
\begin{eqnarray}
z_{ab} &:&\left( 0,0;{\bf 28}_{+}\right) ,\quad z_{\dot{a}\dot{b}
}^{*}:\left( 0,0;{\bf 28}_{-}\right) ,\,\,\rightarrow \quad 56\,\,real \\
\vec{F}_{ab} &:&\left( 1,0;{\bf 35}+{\bf 1}\right) ,\quad \vec{F}_{\dot{a}
\dot{b}}^{*}:\left( 0,1;{\bf 35}+{\bf 1}\right) ,\,\,\rightarrow 216\,\,real
\\
P_{\mu m} &:&\left( 1/2,1/2;{\bf 8}_V\right) ,\quad \rightarrow
\,\,32\,\,real \\
A_{\mu m_1m_2m_3} &:&\left( 1/2,1/2;{\bf 56}_V\right) ,\quad \rightarrow
\,\,224\,\,real
\end{eqnarray}
The operators $P_{\mu m}$ are identical to the operators that come directly
from the reduction from 12D in (\ref{comp}) $P_{\mu m}\equiv Z_{\mu m},$ and
they include the usual momentum $P_\mu \equiv P_{\mu 0^{\prime }}.$ The
Lorentz singlet {\it complex }antisymmetric{\it \ }central extensions $
z_{ab},z_{\dot{a}\dot{b}}^{*},$ classified as the ${\bf 28}_{\pm }$ of $
SO(7,1),$ are directly related to the {\it two real} ${\bf 28}_V$ of $SO(7,1)
$ that are obtained from the reduction from 12D in eq.(\ref{comp}): 
\begin{eqnarray}
e_{mn} &\equiv &Z_{mn},\,\,\,\,\,\,m_{mn}\varepsilon _{\mu _0\mu _1\mu _3\mu
_3}\equiv Z_{mn\mu _0\mu _1\mu _3\mu _3}^{+} \\
z_{ab} &=&\gamma _{ab}^{mn}\,\left( e_{mn}+im_{mn}\right) ,\,\,\,z_{\dot{a}
\dot{b}}^{*}=\gamma _{\dot{a}\dot{b}}^{mn}\,\left( e_{mn}-im_{mn}\right) 
\end{eqnarray}
where the $\gamma _{ab}^{mn},\gamma _{\dot{a}\dot{b}}^{mn}\,$ are
antisymmetric and given by 
\begin{equation}
\gamma ^{0^{\prime }k}=\pm i\gamma ^k,\quad \gamma ^{kl}=\frac i2[\gamma
^k,\gamma ^l].
\end{equation}
These $\gamma _{ab}^{mn},\gamma _{\dot{a}\dot{b}}^{mn}$ satisfy the $SO(7,1)$
Lie algebra in the ${\bf 8}_{\pm }$ representations. The $e_{mn},$ $
m_{mn}\,\,$ have the interpretation of ``electric'' and ``magnetic'' central
charges respectively, as can be gathered from their properties under parity
(the $\varepsilon _{\mu _0\mu _1\mu _3\mu _3}$ is pseudoscalar). The
blackhole entropy is a function of only these Lorentz singlet central
charges. Evidently, both $e_{mn},$\thinspace $m_{mn}\,$are in the adjoint 
{\bf 28} of $SO(7,1)$ \footnote{
Note also that these are not the same as the $q_{ij},p^{ij}$ defined in
other references (e.g. \cite{kal}) since those are in an SO(8) basis rather
than the SO(7,1) basis used here. The relation between the two is not
trivial, and can be obtained from the following discussion.}. 

The remaining operators are not strictly central since they do not commute
with the Lorentz generators (but they commute with each other and with the
supercharges, in flat spacetime). To begin with, the Lorentz vectors $P_{\mu
m}$ are not central. The Lorentz axial vectors $A_{\mu m_1m_2m_3}$ are
related to 12D by 
\begin{equation}
A_{\mu m_1m_2m_3}\varepsilon _{\,\,\,\mu _1\mu _2\mu _3}^\mu \equiv
Z_{m_1m_2m_3\mu _1\mu _2\mu _3}^{+}.
\end{equation}
The complex symmetric $\vec{F}_{ab}$ together with their complex conjugates
can be rewritten in the form of Lorentz antisymmetric tensors ($F_{\mu \nu
})_{ab},$ which correspond to the ({\bf 1+35}) Lorentz-antisymmetric tensors 
$Z_{\mu \nu }$ and $Z_{\mu \nu m_1m_2m_3m_3}^{+}$ that come from the 12D
reduction (\ref{comp}), 
\begin{equation}
(F_{\mu \nu })_{ab}\equiv Z_{\mu \nu }\delta _{ab}+Z_{\mu \nu
m_1m_2m_3m_3}^{+}(\gamma ^{m_1m_2m_3m_3})_{ab},
\end{equation}
where the $SO(7,1)\,\,$ self dual $Z_{\mu \nu m_1m_2m_3m_3}^{+}$ gives the 
{\bf 35} of $SO(7,1).$ Other components of $Z_{M_1\cdots M_6}^{+}$ not
mentioned so far are equivalent to the ones already discussed, because of
its self duality properties. This accounts for the correspondance between
the 528 operators in 4D and 12D notations, and gives a 12D interpretation to
the 4D superalgebra (\ref{4d}). This kind of reclassification of the 528 12D
type-A superalgebra operators (as well as of the 13D type-B superalgebra
operators) in compactifications to each lower dimension was given in a table
in \cite{ibbeyond}. 

As explained in \cite{ibstheory}, Lorentz singlet central extensions
correspond to charges of point particles (or black holes), while Lorentz
non-singlet extensions correspond to boundaries of charged $p$-branes. The
number of anti-symmetrized Lorentz indices corresponds to $p$. Therefore, $
P_{\mu m}$ ($m\neq 0^{\prime }),\,A_{\mu m_1m_2m_3}$ correspond to end
points of strings in 4D, and $\vec{F}_{ab}$ correspond to boundaries of
membranes in 4D. 
\begin{eqnarray}
points &\rightarrow &z_{ab},z_{\dot{a}\dot{b}}^{*} \\
strings &\rightarrow &P_{\mu m}(m\neq 0^{\prime }),\,A_{\mu m_1m_2m_3} \\
membranes &\rightarrow &\vec{F}_{ab},\vec{F}_{\dot{a}\dot{b}}^{*}
\end{eqnarray}

If all the 528 operators in the superalgebra (\ref{4d}) are kept, there
really is a full $SO(10,2)$ isometry in disguise, since this is equivalent
to the superalgebra (\ref{12d}). In this paper we are interested in
describing just the black hole sector for which the algebra is evaluated on
states that satisfy \footnote{
All bosonic operators commute in flat spacetime \cite{ibstheory}, therefore
they can be simultaneously diagonalized.} 
\begin{equation}
black\,\,hole\,\,sector:\vec{F}_{ab}=\vec{F}_{\dot{a}\dot{b}}^{*}=A_{\mu
m_1m_2m_3}=0.  \label{bhsector}
\end{equation}
So far the superalgebra simplifies to 
\begin{eqnarray}
\left\{ Q_{\alpha a},Q_{\beta b}\right\}  &=&\left( i\sigma _2\right)
_{\alpha \beta }\,z_{ab}  \nonumber \\
\left\{ Q_{\dot{\alpha}\dot{a}},Q_{\dot{\beta}\dot{b}}\right\}  &=&\left(
i\sigma _2\right) _{\alpha \beta }\,z_{\dot{a}\dot{b}}^{*}  \label{reduced}
\\
\left\{ Q_{\alpha a},Q_{\dot{\beta}\dot{b}}\right\}  &=&\sigma _{\alpha \dot{
\beta}}^\mu \,\gamma _{a\dot{b}}^m\,P_{\mu m}  \nonumber
\end{eqnarray}
Although the full $SO(10,2)$ isometry is lost by concentrating on only this
sector, there still is an $SO(3,1)\times SO(7,1)$ isometry that has
information on the 12th dimension through $SO(7,1)$ that rotates the 12th
dimension into the other compactified dimensions, including the 11th. This
superalgebra still allows the presence of strings. 

If we also set $P_{\mu m}$ ($m\neq 0^{\prime })$ equal to zero to eliminate
the charged black strings, then using eq.(\ref{gam8}) $\gamma _{a\dot{b}
}^{0^{\prime }}=\delta _{a\dot{b}}$ for Weyl spinors, we obtain the standard
supersymmetry algebra with the usual Lorentz singlet extensions, consistent
with previous discussions of black holes. However, this further restriction
on the $P_{\mu m}$ breaks the $SO(7,1)$ isometry. Instead, we need to
restrict the superalgebra to the black hole sector consistently with $SO(7,1)
$ covariance, which is important to keep track of the 12th dimension. Such a
covariant condition is 
\begin{equation}
P_{\mu m}=P_\mu v_m  \label{product}
\end{equation}
where $P_\mu $ plays the role of 4D momentum and $v_m$ plays the role of 8D
``momentum'' in the compactified directions. Now, since there is $SO(7,1)$
covariance, it is possible to apply a boost to rotate $v_m$ to the form $
v_m=\left( 1,0,\cdots ,0\right) ,$ to work in a $SO(7,1)$ ``rest frame'' in
which the superalgebra takes the standard form $\left\{ Q_{\alpha a},Q_{\dot{
\beta}\dot{b}}\right\} =\sigma _{\alpha \dot{\beta}}^\mu \,\delta _{a\dot{b}
}\,P_\mu .$ Of course, this could be done for one given eigenvalue of the
operator $v_m$, it cannot be done simultaneously for all the states of the
secret theory, unless the whole theory is restricted to a single eigenvalue
of the operator $v_m$. More generally, in the secret theory, we will allow
all eigenvalues of $v_m$ just as we allow all eigenvalues of $P_\mu .$
Evidently, now there is a parallel between the 4D and 8D sectors.

\subsubsection{Map between $SO(7,1)$ and $SU(8)$ classifications}

If one uses the product form (\ref{product}) one cannot see the
classification of the operators under $K\subset U.$ The $SU(8)$ structure of
eq.(\ref{4dz}) is not evident in the $SO(7,1)$ covariant form 
\begin{equation}
\left\{ Q_{\alpha a},Q_{\dot{\beta}\dot{b}}\right\} =\sigma _{\alpha \dot{
\beta}}^\mu \,P_\mu \,\,\gamma _{a\dot{b}}^mv_m.\,  \label{so71}
\end{equation}
An $SU(8)$ classification is not possible unless $v_m$ is boosted with an $
SO(7,1)$ transformation to the ``rest frame'' $v_m=\left( 1,0,\cdots
,0\right) $ (we have normalized $v^mv_m=1$ by absorbing the overall scale
into $P_\mu $). This kind of phenomena has been noticed before in another
context: in order to display electric-magnetic duality as a symmetry, one
has to give up a Lorentz covariant formalism. Similarly, in the present
case, in order to display a form consistent with $SU(8)$ (i.e. duality) we
have to give up an $SO(7,1)$ covariant formalism, or vice versa. 

This remark
leads to the map $T$ between the $SU(8)$ basis and the $SO(7,1)$ spinor
basis. Consider a map $T$ that relates the two bases 
\begin{equation}
Q_{\alpha A}=T_A^{\,\,\,\,a}Q_{\alpha a},\quad Q_{\dot{\alpha}\bar{A}}=T_{
\bar{A}}^{\,\,\,\,\dot{a}}Q_{\dot{\alpha}\dot{a}},\quad T_{\bar{A}}^{\,\,\,\,
\dot{a}}=\left( T_A^{\,\,\,\,a}\right) ^{*}
\end{equation}
where $A=1,\cdots ,8$ denotes the ${\bf 8}$ of $SU(8).$ When applied to (\ref
{so71}) it gives the desired $SU(8)$ covariant result 
\begin{equation}
\left\{ Q_{\alpha A},Q_{\dot{\beta}\bar{B}}\right\} =\sigma _{\alpha \dot{
\beta}}^\mu \,P_\mu \,\,\delta _{A\bar{B}},\quad \Leftrightarrow \,\,\left[
T\left( \gamma ^mv_m\right) T^{\dagger }\right] _{A\bar{B}}=\delta _{A\bar{B}
}
\end{equation}
$T_A^{\,\,\,\,a}(v)$ is just the $SO(7,1)$ boost (in the spinor
representation) and it can be constructed explicitly 
\begin{equation}
T=\frac{1+v_{0^{\prime }}-\vec{\gamma}\cdot \vec{v}}{\sqrt{2+2v_{0^{\prime }}
}}=T^{\dagger },\quad T^{-1}=T^T=T^{*}
\end{equation}
Here $\vec{\gamma}\cdot \vec{v}$ contains only the seven hermitian purely
imaginary gamma matrices. $T^{-1}$ is just the square root of the 8D
``momentum'' matrix $\gamma ^mv_m$ since it gives 
\begin{equation}
\left( T^{-1}\right) ^2=v_{0^{\prime }}+\vec{\gamma}\cdot \vec{v}=\gamma
^mv_m,\quad {\rm {with\,\,\,\,}}v_{0^{\prime }}^2-\vec{v}^2=1.{\rm {\,}}
\label{square}
\end{equation}
The $SU(8)$ basis central extensions that appear in (\ref{4dz}) are now
written as 
\begin{equation}
Z_{AB}=(TzT^T)_{AB},\quad \quad Z_{\bar{A}\bar{B}}^{*}=(T^{*}z^{*}T^{T*})_{
\bar{A}\bar{B}}=T^Tz^{*}T.  \label{t4}
\end{equation}
Hence the central extensions $z_{ab},Z_{AB}$ are functions of not only the
moduli $E_{7\left( 7\right) }/SU(8)$ but also of the additional moduli $
v_m\sim SO(c+1,1)/SO(c+1).$ We will use this fact in section 4 in order to
show the invariance of the entropy.

\subsection{Superalgebra in 5D}

The story in 5D is similar, but has an additional important point which is
not present in the 4D superalgebra. This consists of one additional Lorentz
scalar central extension that is needed for $SO(c+1,1)=SO(6,1)$ covariance,
but which is absent in the $K=Sp(8)$ basis used in eqs.(\ref{5dz},\ref{ent5}
). Understanding this point brings more clarity to the 12D point of view. In
the remaining dimensions there are no new issues, hence they will not be
discussed here.

Following the same strategy that gave (\ref{reduced}) and (\ref{product}),
we obtain the $d=5,\,c=5$ superalgebra by compactification from 12D, keeping
only the relevant terms for the black hole sector 
\begin{eqnarray}
\left\{ Q_{\alpha a},Q_{\beta b}\right\}  &=&\left( C\gamma ^\mu \right)
_{\alpha \beta }\left( \gamma ^m\right) _{ab}P_{\mu m}+C_{\alpha \beta
}\,\,\,y_{ab}  \label{so61} \\
P_{\mu m} &=&P_\mu v_m,\quad y_{ab}=\left( e_{mn}\gamma ^{mn}+m_m\gamma
^m\right) _{ab}  \nonumber
\end{eqnarray}
where $a$ denotes the 8-spinor index of $SO(6,1)$ and $m=0^{\prime
},1,\cdots ,6$ is its vector index. The 7 gamma matrices $\gamma _{ab}^m$
are antisymmetric \footnote{
The seven 8$\times 8$ gamma matrices that appear in eq. (\ref{gam8}) have
been relabelled into $m=0^{\prime },1,\cdots ,6,$ and by multiplying one of
them with $i.$ Note that $\gamma ^{0^{\prime }}$ is not proportional to 1 in
the $SO(6,1)$ Dirac spinor space.}. The 21 $\gamma ^{mn}$ are also
antisymmetric matrices and form the $SO(6,1)$ Lie algebra. The relation of
the ``electric'' and ``magnetic'' central extensions $e_{mn},\ m_m$ to the
12D tensors in (\ref{12d}) is given by 
\begin{equation}
e_{mn}\equiv Z_{mn},\quad m_m\,\varepsilon _{\mu _0\mu _1\cdots \mu
_4}=Z_{m\mu _0\mu _1\cdots \mu _4}^{+}.
\end{equation}

The new issue is that the $SO(6,1)$ covariant form (\ref{so61}) requires
7+21=28 central extensions, while the $Sp(8)$ covariant form (\ref{5dz}) has
only 27 of them. Although the structure of the superalgebra permits one more
central extension that is an $Sp(8)$ singlet, this central extension does
not appear in the known 5D supergravity theories that come from
compactifications of 11D supergravity (see e.g. \cite{cremmer}). 

To relate the $SO(6,1)$ spinor basis to the $Sp(8)$ basis we may again
construct the transformation that corresponds to a $SO(6,1)$ boost 
\begin{eqnarray}
Q_{\alpha A} &=&T_A^{\,\,\,\,\,a}Q_{\alpha a},\quad T\left( v\cdot \gamma
\right) T^T=\Omega \equiv \gamma ^{0^{\prime }} \\
T &=&\frac{1+v_{0^{\prime }}-\gamma _{0^{\prime }}\vec{\gamma}\cdot \vec{v}}{
\sqrt{2+2v_{0^{\prime }}}},\,\,\,\,\,\,T^{-1}=T^T
\end{eqnarray}
Then the relation between the central extensions in the two bases is 
\begin{equation}
y=T^TYT=e_{mn}\gamma ^{mn}+m_m\gamma ^m  \label{t5}
\end{equation}
Since $Y_{AB}$ is orthogonal to $\Omega _{AB},$ this implies that 
\begin{equation}
m\cdot v=0.
\end{equation}
This condition is $SO(6,1)$ invariant, and it can be imposed only in the
presence of the additional quantum numbers $v_m$ (perhaps this last
restriction may be removed in a more general supergravity theory of the type
suggested in the last section). 

We now see that the 5D superalgebra can be rewritten covariantly under $
SO(6,1)$, and that the central extensions depend not only on the moduli $
E_{6\left( 6\right) }/Sp(8)$ but also on the moduli $v_m$ described by $
SO(c+1,1)/SO(c+1).\,\,$The additional moduli play a crucial role in
understanding the invariance of the entropy and of the hidden 12th dimension.

\section{Invariance of the mass and entropy}

We are now ready to consider the extreme black holes. These corespond to
short representations of the superalgebra since they satisfy the BPS
conditions (the discussion is similar for the nearly extreme ones). In the
S-theory language this corresponds to requiring that the determinant of the
32$\times 32$ matrix on the right hand side of the 12D superalgebra (\ref
{12d}) vanishes on such states 
\begin{equation}
det\left( S\right) \,|BPS>=0.
\end{equation}
This condition is equivalent to the vanishing of some linear combination of
supercharges on the BPS states. This gives short multiplets of the
superalgebra, corresponding to the BPS states. If the multiplicity of the
zero eigenvalue of $S$ is $2n,$ then the size of the short supermultiplet is 
\begin{equation}
(2_B^{15-n}+2_F^{15-n})\times dimR  \label{dim}
\end{equation}
where $(2_B^{15-n}+2_F^{15-n})$ corresponds to the number of independent
bosonic (B) or fermionic (F) terms in the most general polynomial
constructed with the supergenerators, in the sector with $2n$ zero
eigenvalues. 

The factor $dim(R)$ corresponds to the (collection of) representations $R$
of the reference state on which the supergenerators are applied. These {\it 
must correspond to a representation of the isometries of the algebra}, which
we will call the ``little group'' in analogy to the little group in the
representation theory of the (super)Poincar\'e group. The little group in our
case includes either $K$ or $C$. The specific representations $R$ that are
relevant depend on the details of the secret theory, but one can try to
extract this information from one of its limits, such as string theory and
its generalizations (including D-branes), by combining that information with
consistency with the little group. 

This representation theory method of
counting states \cite{ib11} will not be applied in the present paper.
Instead, we will use the already known results given in the introduction,
and concentrate on the additional invariances that are expected to be
present under the little group, because of the procedure outlined above. By
showing that the entropy, computed with other methods, is invariant under
the little group, we build up evidence that the global structure of S-theory
(together with its implied extra dimensions) is correct.

The condition $det\left( S\right) =0$ is invariant under all of the
isometries, $K$ or $C$ in every dimension, applied on $S.$ In the black hole
sector of eqs.(\ref{bhsector},\ref{product}) there still is $K$ or $C$
isometry depending on the $SO(c+1,1)$ frame chosen. $det\left( S\right) $
remains invariant under the boost transformations $T$  that relates the two
bases. Therefore, one has the same set of solutions $|BPS>$ in either the $K$
or $C$ bases, but the states are regrouped differently into a collection of
representations $R$ of either $K$ or $C$. That is, the representations $R$
are the same collection of states, that can be expanded as either
representations of $K$or of $C.$ Since the counting of the states, $dimR,$
is the same in any frame, it must be true that the entropy is invariant
under both $K$ and $C$. 

In the general $SO(c+1,1)$ frame the determinant condition takes the form 
\begin{equation}
4D:\quad det\left( 
\begin{array}{cc}
i\sigma _2\otimes z & \sigma ^\mu P_\mu \otimes \gamma ^mv_m \\ 
\sigma ^{T\mu }P_\mu \otimes \gamma ^{Tm}v_m & i\sigma _2\otimes z^{*}
\end{array}
\right) =0
\end{equation}
\begin{equation}
5D:\quad det\left( C\gamma ^\mu P_\mu \otimes \gamma ^mv_m+C\otimes y\right)
=0
\end{equation}
These can be simplified (the normalization $v^2=v_{0^{\prime }}^2-\vec{v}^2=1
$ is used without loss of generality by absorbing the norm into the
definition of $P_\mu )$ 
\begin{eqnarray}
4D &:&\quad \left[ det\left( P^21_8-vzv^*z^{*}\right) \right] ^2=0, \\
5D &:&\quad \left[ det\left( P^21_8-vyvy\right) \right] ^2=0.
\end{eqnarray}
Therefore in 4D (or 5D) the black hole mass $P^2=M^2$ is given by the
largest eigenvalue of the 8$\times 8$ matrix $vzv^Tz^{*}$ (or $vyvy).$ The
eigenvalues are obviously $SO(c+1,1)$ invariants. By using this invariance
we can apply a boost to go to the frame $v_m\rightarrow (1,\vec{0}$ ) to
show that in 4D (or 5D) the eigenvalue is the same as the one obtained by
diagonalizing ($ZZ^{*})_{A\bar{B}}$ in the $SU(8)$ basis (or $\left( Y\Omega
Y\Omega \right) _{AB}$ in the $Sp(8)$ basis). Therefore, as expected, the
mass is both $C$ as well $K$ invariant.

Turning to the entropy, we use the map $T^{-1}$ in eq.(\ref{t4},\ref{t5})
from the $K$ basis to the $C$ basis, and recall eq.(\ref{square}) $\left(
T^{-1}\right) ^2=v,$ to obtain new expressions for the entropy in 4D and 5D,
in terms of the invariants 
\begin{eqnarray}
I_4 &=&Tr\left( vzv^{*}z^{*}vzv^{*}z^{*}\right) -\frac 14\left[ Tr\left(
vzv^{*}z^{*}\right) \right] ^2+4Pf\left( z\right) +4Pf\left( z^{*}\right) ,
\\
I_5 &=&Tr\left( vyvyvy\right) 
\end{eqnarray}
These are equivalent to the ones given in the introduction, but are now
written explicitly as invariants of $SO(c+1,1)$ that mixes the 12th (or
13th) dimension with the other compactified dimensions. In the ``rest
frame'' $v=(1,\vec{0})$ the new form reduces to the form given in the
introduction, after using $\gamma ^{0^{\prime }}=1$ for $SO(7,1)$, or $
\gamma ^{0^{\prime }}=\Omega $ for $SO(6,1)$. The entropy really depends
only on the quantized quantum numbers, but they are written in terms of
central extensions that depend not only on the moduli $U/K$ but also on the
moduli $v_m\sim SO(c+1,1)/SO(c+1).$

\section{Future directions}

We have seen that a 12D interpretation of the black hole entropy is
possible. This provides some evidence for the existence of a 12th dimension
that is apparently hidden in string theory black holes. It must be
emphasized that the discussion was given for a 12D description of the A-type
supersymmetry. By T-duality in S-theory, going over to the B-type
superalgebra, this is equivalent to finding the 13th dimension in the
blackhole entropy.

There are several directions for expanding on the remarks made in this paper.

1) We have seen that the superalgebra is covariant under either $K$ or $C,$
but it cannot be covariant simultaneously under both. This should have been
expected by analogy to the incompatibility of Lorentz covariance with
electric-magnetic duality. It is known that these cannot be displayed as
simultaneous covariances of the equations. In the present case $K$ is
related to dualities and $C$ to Lorentz transformations. However, one
quantity, the entropy, is invariant under both $C$ and $U.$ Since the former
is generally not fully included in $U$ (see footnote 6), except for the
subgroup $SO(c+1)$, it must be that there is a larger group of symmetries $G$
that leave the entropy invariant, such that 
\begin{eqnarray}
G &\supseteq &U\supset K,\quad and\;\,\,\,G\supset SO(c+1,1) \\
K &\supset &SO(c+1) \subset SO(c+1,1).
\end{eqnarray}
This group, which includes duality transformations as well as rotations into
the hidden dimensions, is likely to provide valuable information about the
secret theory. Work in progress on this point will be reported elsewhere.

2) In order to have a covariant description including the 12th ``dimension''
we needed the form $P_{\mu m}=P_\mu v_m$ of eq.(\ref{product}), which
implies the existence of extensions $P_{\mu m}\,\,(m\neq 0^{\prime })$ in
the superalgebra that are Lorentz non-singlets. As already emphasized, this
requires the presence of open strings, hence the presence of non-locality
which is natural in a theory of extended objects.

Another way of understanding the non-locality is as follows. In the sector
in which $P_{\mu m}=$ $P_\mu v_m$ the states are labelled by the eigenvalues
of both operators $|P_\mu $,$v_m>.$ In a field theory description one will
need to introduce bi-local fields in the Fourier space $\Phi (x_\mu ,y_m),$
consistent with non-locality. The Kaluza-Klein expansion in the $y$ variable
gives a tower of local fields $\Phi _{v_m}(x_\mu )$ labelled by the
eigenvalues of $v_m,$ all of which have degenerate mass $P^2v^2=M^2.$ We
have seen that the black hole entropy and mass is consistent with the
extension of spacetime in this sense.

``Spacetime'' begins to have a new meaning, and we begin to enter an unknown
territory.

3) More generally, according to S-theory, in the presence of open p-branes,
the superalgebra should include a larger set of Lorentz non-singlet
extensions. In their presence the concept of spacetime is expanded even
further than above. In this context, our usual spacetime, as represented by
the usual momentum $P_\mu $ in the superalgebra, does not seem to be more
special as compared to the other extensions. In our current level of
understanding the new parts of ``spacetime'' are interpreted as open
p-branes. In the future we may find that the central extensions suggested in
S-theory may take a more democratic role as parts of an expanded, much
richer, ``spacetime''.

4) It would be interesting to construct models of extended objects that fit
this description of spacetime. One possibility is to consider their low
energy effective theory in which one keeps some of their degrees of freedom,
as in the example of a bi-local field above. A guiding principle is that the
physical states should fall into a representation of the superalgebra $
\left\{ Q,Q\right\} =S,$ with $detS=0,$ which is a condition for
representations that correspond to both massless and massive BPS states. 

Such an example in 12D was given in \cite{ibstheory} in the form $S_{\alpha
\beta }=\gamma _{\alpha \beta }^{\mu \nu }(p_\mu v_\nu -p_\nu v_\mu ),$ for
which the determinant condition becomes $p^2v^2-(p\cdot v)^2=0.$ In this
case $S$ has 16 zero eigenvalues, therefore, according to eq.(\ref{dim}),
the number of states is $2_B^7+2_F^7$ for the smallest multiplet. This
representation is obviously covariant under SO(10,2) by construction, but
yet it contains a set of fields that is identical to the one in 11D
supergravity. Furthermore, in the present case the fields are bi-local $\Phi
(x_\mu ,y_\nu )$, and the variables are 12 dimensional with signature $
\left( 10,2\right) $. This suggests that a 12D supergravity that corresponds
to this representation should exist. A plan for constructing it could
include using the old gauging principles that were applied in the
construction of usual supergravity. The global algebra that provides the
starting point is the superalgebra of S-theory. 

The connection to the usual 11D supergravity can be seen as follows. The
determinant constraint can be satisfied by taking equations of motion that
correspond to $p^2=0,$ $p\cdot v=0$ such as 
\begin{equation}
\partial _x^2\Phi (x_\mu ,y_\nu )+\cdots =0,\quad \partial _x\cdot \partial
_y\Phi (x_\mu ,y_\nu )+\cdots =0  \label{12dsugra}
\end{equation}
where the $\cdots $ represent non-linear interactions. If one takes a
Kaluza-Klein expansion in the $y$ variable one has the tower $\Phi _{v_\mu
}(x_\mu )$ labelled by $v_\mu .$ By keeping only one of these eigenvalues,
e.g. $v_\mu =(1,0,\cdots ,0)$ the equations would reduce identically to 11D
supergravity.

5) It is worthwhile to try to build 12D models as the candidates for the low
energy physics of the secret theory, and to shed more light on its
structure. There has been some work involving SO(10,2) models that use the
specialized superalgebra $S_{\alpha \beta }=\gamma _{\alpha \beta }^{\mu
\nu}(p_\mu v_\nu -p_\nu v_\mu )$ suggested in \cite{ibstheory}. These are
super Yang-Mills type fields \cite{seznish} and strings \cite{martinec} that
fit into the S-theory framework. However, these constructions were
restricted to a single eigenvalue of $v_\mu ,$ and therefore they break the $
SO(10,2)$ symmetry. Their generalizations in the sense of (\ref{12dsugra})
would remove this restriction.

6) It would be interesting to try to understand the $SO(c+1,1)$ symmetry
discussed here from the point of view of M- and F-theories \cite{witten}- 
\cite{vafaf}.

\vfill\eject


\begin{thebibliography}{99}
\bibitem{st-vaf}  A. Strominger and K. Vafa, Phys.Lett.{\bf B379}(1996) 104
[= hep-th/9601029 ]; J. Breckenridge, D. Lowe, R. Myers, A. Peet, A.
Strominger, C. Vafa, Phys. Rev. Lett. {\bf B381} (1996) 423 [hep-th/9603078].

\bibitem{horstr}  G. Horowitz and A. Strominger, Phys.Lett. {\bf 77} (1996)
2363 (hep-th/9602051); G. Horowitz, J. Maldacena, A. Strominger, Phys. Lett. 
{\bf B383} (1996) 151 [hep-th/9603109].

\bibitem{hor}  G.Horowitz, `` The origin of black hole entropy in string
theory", gr-qc/9604051.

\bibitem{cvet}  M. Cvetic and D. Youm, Phys.Rev.{\bf D54} (1996) 2612 [=
hep-th/9603147], hep-th/9605051, hep-th/9603100.

\bibitem{cvetT}  M. Cvetic and A.A. Tseytlin, hep-th/9510097, hep-th/9512031.

\bibitem{suss}  E. Halyo, B. Kol, A. Rajaraman, L. Susskind, ``Counting
Schwarzchild and charged black holes", hep-th/9609075.

\bibitem{ver}  R. Dijkgraaf, E. Verlinde, H. Verlinde, ``BPS spectrum of the
five brane and black hole entropy", hep-th/9603126 ; ``BPS quantization of
the five brane", hep-th/9604055.

\bibitem{kal}  R. Kallosh and B. Kol, Phys.Rev. D53 (1996) 5344 [=
hep-th/9602014].

\bibitem{kal-fer}  S. Ferrara and R. Kallosh, Phys.Rev. D54 (1996) 1514 [=
hep-th/9602136], Phys.Rev. D54 (1996) 1525 [= hep-th/9603090]

\bibitem{hc}  M. Cvetic and C. Hull, ``Black holes and U duality",
hep-th/9606193.

\bibitem{bkal}  K. Behrndt, R. Kallosh, J. Rahmfeld, M. Shmakova, and W. Kai
Wong, ``STU black holes and string triality", hep-th/9608059.

\bibitem{kal-raj}  R. Kallosh and A. Rajaraman, ``Brane-antibrane
democracy'', {hep-th/9604193}.

\bibitem{pol}  J. Polchinski, Phys.Rev.Lett. {\bf 75} (1995) 4724
(=hep-th/9510017). See also, J. Polchinski, S. Chaudhuri, C.V. Johnson
``Noteson D-branes'', hep-th/9602052.

\bibitem{ht}  C. Hull and P. Townsend, Nucl.Phys.{\bf B438} (1995) 109 [=
hep-th/9410167].

\bibitem{ibbeyond}  I. Bars, ``Supersymmetry, p-brane duality and hidden
spacetime dimensions'', Phys. Rev. {\bf D54} (1996) 5203 [= hep-th/9604139];
``Duality and hidden dimensions'', hep-th/9604200, to appear in the
proceedings of Frontiers in Quantum Field Theory, Toyonaka, Japan, Dec. 1996.

\bibitem{ibstheory}  I. Bars, ``S-Theory'', hep-th/9607112; ``Algebraic
Structures in S-Theory'', hep-th/9608061.

\bibitem{witten}  E. Witten, Nucl. Phys. B443 (1995) 85; and ``Some comment
son String Dynamics'', hep-th/9507121, to appear in the proc. of Strings
'95; P. Horava and E. Witten, Nucl. Phys. {\bf B460 (}1996) 506
[=hep-th/9510209].

\bibitem{townsend}  P. Townsend, ``p-brane democracy'', hep-th/9507048.

\bibitem{jhs}  J. Schwarz, Phys.Lett.{\bf B367} (1996) 97 [=hep-th/9510086];
hep-th/9509148; ``M-theory extensions of T duality'', hep-th/9601077.

\bibitem{vafaf}  C. Vafa, Nucl.Phys. {\bf B469} (1996) 403
[=hep-th/9602022]; D.R. Morrison and C. Vafa, Nucl.Phys. {\bf B473} (1996)
74 [= hep-th/9602114], hep-th/9603161; E. Witten, Nucl.Phys. {\bf B471}
(1996) 195 [=hep-th/9603150].

\bibitem{ib11}  I. Bars, Phys. Rev. {\bf D52} (1995) 3567 [=hep-th/9503228];
I. Bars and S. Yankielowicz, Phys. Rev. {\bf D53} (1996) 4489
[=hep-th/9511098]. I. Bars, ``Consistency between 11D and
U-duality'',hep-th/9601164, in Ahrenshoop Symp.1995:115-122 (
QCD161:S937:1995 ).

\bibitem{cremmer}  E. Cremmer, ``On hidden symmetries in extended
supergravities" LPTENS 83/1, Jan 1983. Lectures given at September School on
Supergravity and Supersymmetry, Trieste, Italy, 1982, Published in Trieste
Workshop 1982:153 ( QC178:T7:1982 )

\bibitem{seznish}  H. Nishino and E. Sezgin, ``Supersymmetric Yang-Mills
equations in (10+2)-dimensions", hep-th/9607185.

\bibitem{martinec}  E. Martinec, Geometrical structures of M-theory,
hep-th/9608017.
\end{thebibliography}
\end{document}